\pdfoutput=1
\documentclass{acm_proc_article-sp}
\usepackage{graphicx}
\usepackage{subfigure}
\title{Finding Nemo: Searching and Resolving Identities of Users Across Online Social Networks}
\author{Paridhi Jain, Ponnurangam Kumaraguru \\ Indraprastha Institute of Information Technology (IIIT-Delhi), India \\ \{paridhij, pk\}@iiitd.ac.in \\ \large{precog.iiitd.edu.in}}

\begin{document}
\maketitle
\begin{abstract}
An online user joins multiple social networks in order to enjoy different services. On each joined social network, she creates an identity and constitutes its three major dimensions namely profile, content and connection network. She largely governs her identity formulation on any social network and therefore can manipulate multiple aspects of it. With no global identifier to mark her presence uniquely in the online domain, her online identities remain unlinked, isolated and difficult to search. Earlier research has explored the above mentioned dimensions, to search and link her multiple identities with an assumption that the considered dimensions have been least disturbed across her identities. However, majority of the approaches are restricted to exploitation of one or two dimensions. We make a first attempt to deploy an integrated system \emph{Finding Nemo} which uses all the three dimensions of an identity to search for a user on multiple social networks. The system exploits a known identity on one social network to search for her identities on other social networks. We test our system on two most popular and distinct social networks -- Twitter and Facebook. We show that the integrated system gives better accuracy than the individual algorithms. We report experimental findings in the paper. 
\end{abstract}
\section{Introduction}\label{Introduction}
Over the last decade, Online Social Media has evolved from being a platform for broadcasting services (e.g. news, blogs) to a rich multimedia platform for maintaining social connections, thereby introducing online social networks. Each online social network offers a service as their main unique selling proposition e.g. Flickr offers photo sharing, Youtube offers video sharing, Twitter offers micro-blogging. Variety of services offered by multiple online social networks facilitate various ways of information sharing, leading to the massive popularity of social networks.  Users spend more time on Facebook than any other web brand / site,~\footnote{http://blog.nielsen.com/nielsenwire/social/} Twitter claims 600,000 new users and 200 million tweets per day,~\footnote{http://searchenginewatch.com/article/2095216/} and Facebook has 901 million users as of March 2012. The statistics explain the impact of social media on an online user.\\
\indent Owing to a large number of online social networks, popular in different parts of the world, and facilitating different services, a user becomes a member of multiple social networks. A user's presence on multiple social networks helps her to control the nature and audience of information shared (e.g. professional and personal network), and the rate of information dissemination (e.g. viral or restricted)~\cite{Motoyama}. On each joined social network, she creates an identity and constitutes its three major dimensions namely Profile, Content and Network. Each dimension is composed of a set of attributes which describes her and differentiate her from others. Profile is composed of set of attributes which describes her persona such as username, name, age, location etc. Content is composed of attributes which describes the content she creates or is shared with her such as text, time of post etc., and Network is composed of connection attributes which describes the network, she creates to connect to other users such as number of friends. Her identity creation process on each social network involves her significant control and she can choose to give / hide / skip / change her identity attributes, implying that she can constitute any dimension in any form on any identity. Therefore, her multiple identities created by the similar processes, may vary largely. Further, with no handle / identifier / attribute for a user to mark her presence uniquely in Online Social Media domain, her identities remain unlinked and isolated. Because of varied and non-linked identities, it is difficult to find identities of a user on multiple social networks when searched with limited information about her and within a large pool of users identified with the similar information. \\
\indent Literature lists methods to exploit the three dimensions of an identity to enhance information about a user and to gain better hints towards her identities on multiple social networks. Researchers have developed a set of approaches which assume that the considered dimension is constituted in a similar fashion by a user across her multiple identities. Therefore, for that dimension, the attributes hold the same or similar value across her identities, which can be further used to link identities. We term such attributes as \emph{linking attributes}. For example, two identities are linked if profile attributes as name, school and email hold same or similar value on both identities~\cite{Motoyama}; for this illustration, name, school and email are linking attributes. To the best of our knowledge, majority of the approaches proposed exploited either one or two dimensions for an identity search and linking, thereby leaving other hints uninvestigated. We make a first attempt to use all the three dimensions for identity search and linking. We discuss a set of algorithms, one for each dimension, to leverage available information about the user and create a set of candidate identities for a user on a social network. Each algorithm is re-engineered from existing algorithms in the literature to adapt to real-time search, limited availability of information and usage of the auxiliary information left unexplored. We further merge the algorithms to build a system \emph{Finding Nemo} which is capable of exploiting any linking attribute present in any dimension of an identity of a user (details in Section~\ref{Evaluation}). We test our algorithms and the system on two most popular and distinct social networks -- Twitter and Facebook. We evaluate each algorithm and our system on three parameters namely accuracy, candidate set size and search time. Our major contributions are --
\vspace{-2mm}
\begin{itemize}
\item
\vspace{-2mm}
To the best of our knowledge, we deploy the first research based identity search system which exploits three major dimensions of a user's identity namely profile, content and network, together.
\item
\vspace{-2mm}
We use publicly accessible attributes and information to make decisions on identity linking and therefore are independent from any user authorization to gain required information.
\vspace{-2mm}
\item 
We show that users unconsciously and indirectly leak their Facebook identity on Twitter by mentioning web URLs containing their Facebook identity as well as by using third party applications to post content simultaneously on multiple networks.
\vspace{-3mm}
\end{itemize}
%\item
%We search for 52 malicious user's identity on Facebook and could find correct Facebook identity for 27 users, reaching to an accuracy of 52\%. 
\indent The proposed algorithms to find a user's multiple identities can be applied to varied domains. In security domain, our solutions can help searching for malicious user's multiple online identities. Malicious users exploit online social networks to enhance reachability to users (victims). To identify malicious users, security researchers have devised features on individual networks \cite{Grier, Lee, benevenuto@ceas10, Chu, Fabricio, Irani2}. Solutions suggested to detect malicious user accounts are network dependent, hence security analysts need to identify malicious accounts on each networking site. In order to save identification cost and efforts, linking malicious user identities present on multiple online social networks is suggested. Our algorithms capturing behavioral characteristics can help in linking malicious user identity on multiple  social networks. In privacy domain, each algorithm finds its application in understanding the quantity and quality of the user's information leaks. Easily linkable multiple identities of a user can leak a user's private attributes via aggregation of user's information from multiple social networks~\cite{Zheleva1, Krishnamurthy, Goga, Chen}. System analysts can understand the possible privacy leaks and can then improve privacy policies and anonymization methods to preserve user's privacy. We discuss some of the information leaks via public information and aggregation of information in Section~\ref{Discussion}. Our solutions can help in building recommendation feature for social aggregation sites. The recommendation feature can find a user's presence on multiple networks with the given user information on one network and suggest her to aggregate the suggested identities (probably belonging to her), hereby saving user effort. In marketing domain, the algorithms can be applied to aggregate customer data from multiple online shopping sites, to understand customer preferences and buying patterns and then recommend better deals to promote products.\\
\indent The paper is organized as follows;  Section~\ref{Related} presents related literature and motivates the work described in this paper. Section~\ref{Methodology} discusses the proposed algorithms in detail, followed by Section~\ref{Evaluation} which elaborates the evaluation of the algorithms, and motivates the integration of the algorithms to build a better system. We discuss applicability, limitations and other observations in Section~\ref{Discussion} and conclude the paper with future research directions in Section~\ref{Conclusion}. 
% In e-commerce domain, by monitoring online social network identities who have established link with e-commerce site identities, one can understand the stock market impression on users correlate the ups and down in the stock market with their social behavior.
\vspace{-2mm}
\section{Related work} \label{Related}
\begin{figure}[ht]
  \centering
  \includegraphics[ page=1,scale=0.4]{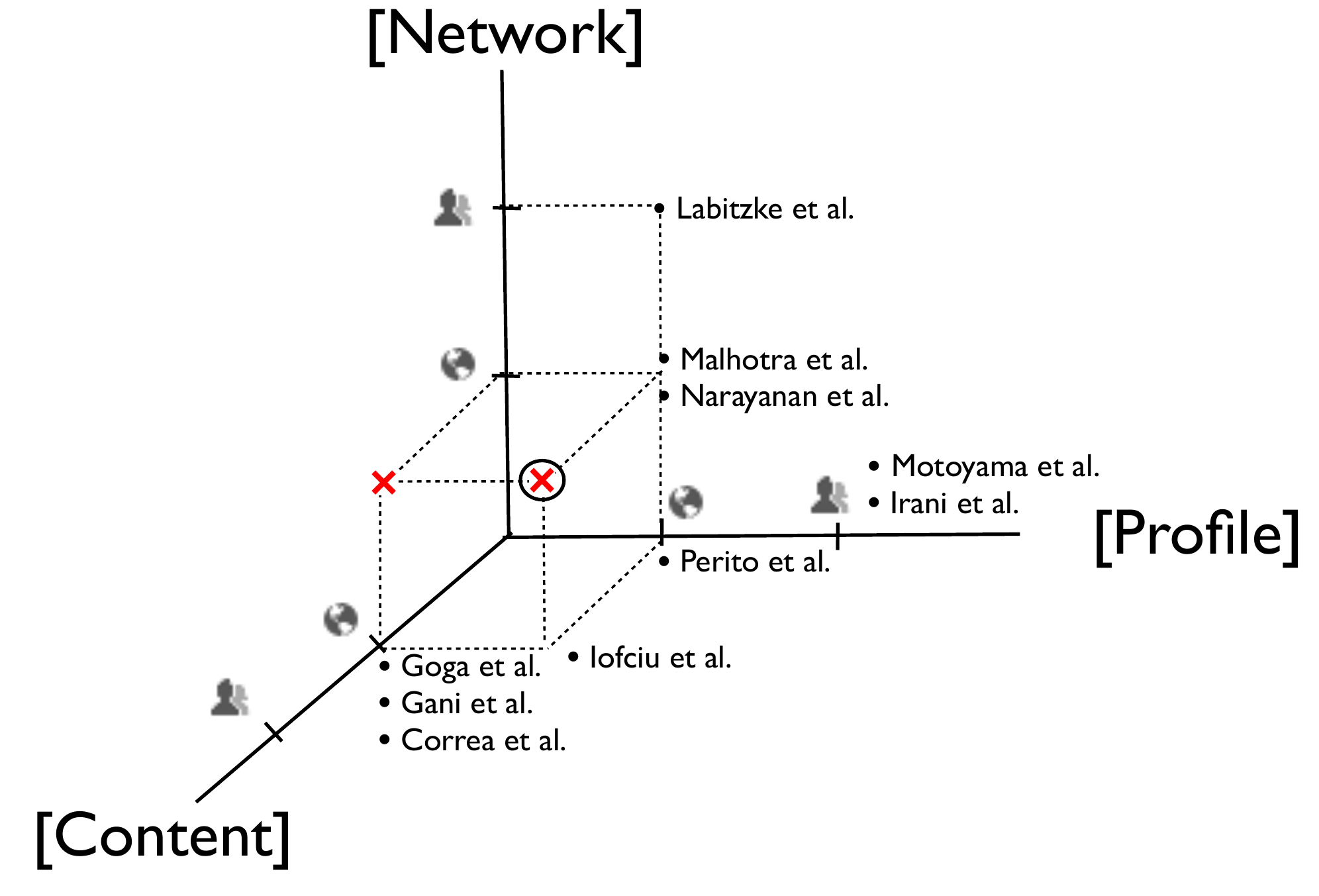}
  \vspace{-3mm}
  \caption{Conceptualization of state-of-the-art identity search and linking approaches across three dimensions of a user's online identity. }
  \label{fig:related}
\end{figure}

Figure~\ref{fig:related} presents a three dimensional conceptualization of state of the art work discussed till date (to the best of our knowledge) for identity search and linking in online social media. Each dimension in the figure represents a distinct aspect of a user's online identity namely Profile, Content and Network as discussed in Section~\ref{Introduction}. Each dimension is further marked with two labels. The outer label represents that the approach discussed in the work accessed private and public attributes, while the inner label represents that the work accessed limited and public attributes.~\footnote{Publicly accessible attributes may vary depending on the social network involved. } We now discuss each of the work in detail.\\
\indent Perito \textit{et al.}, Motoyama \textit{et al.}, and Irani \textit{et al.} suggested methods which use only profile attributes to search and link identities of a user across networks. Perito \textit{et al.} introduced the idea of using public attribute i.e. username to link multiple online identities of a user. The authors suggested to link Google and Ebay identities of a user via extracting string based features from username of identities on two networks. With two features and a dataset of 10,000 username pairs for both positive and negative dataset, they achieved 71\% accuracy~\cite{perito}. On the other hand, Motoyama \textit{et al.} extended the linking attribute set and suggested methods to find and link users' online identities using both private and public attributes. They used name, city, school, location, age, email etc. as linking attributes and used bag-of-words model to match profile attributes on two identities.  With a dataset of 300 profiles for positive and 200 profiles for negative dataset, they achieved 72\% accuracy. They concluded `name', `name $+$ school' and email as strong identity linkers~\cite{Motoyama}.  Irani \textit{et al.} also worked on a set of both private and public attributes, however only categorical attributes and used text matching approaches to match two identities~\cite{Irani}. They concluded that user identities could be accurately linked via a different set of attributes as last name, birth year and country.  \\
 \indent  Owing to the limited set of profile attributes which are public, accessible and available across multiple networks, researchers attempted hybrid approaches to use more than one dimension for identity linking. Malhotra \textit{et al.} exploited profile and network dimension of a user's identity by adding number of friend connections along with public profile attributes to link Twitter and LinkedIn identity of a user. They used attribute specific comparison metrics e.g. Jaro distance for  username and name comparison, Tf-idf vector space model for description comparison etc., to link two identities. The authors evaluated their method over a dataset of 29,129 users, and achieved an accuracy of 98\% with Naive Bayes classification. They proved username and name as the best identity resolvers~\cite{Malhotra}. Iofciu \textit{et al.}, on the other hand, merged public profile attributes with content features and tried to link user identities across social tagging systems using username and tags (content). The authors evaluated their method on three social networks, Flickr, Delicious and StumbleUpon and achieved an accuracy of 60\%. Furthermore, as suggested by Irani \textit{et al.}, the authors also implemented the concept of social footprint. To search for a user's identity on a third network, they aggregated the two linked identities and used the collective information to link the third identity with the social footprint of the user. When evaluated on the same dataset, the authors claimed to link 80\% of the user identities across three social networks correctly~\cite{Iofciu}. \\ 
\indent A set of researchers experimented only with content dimension for identity linking. Goga \textit{et al.} used three attributes extracted from the content created by a user -- timestamp, location and description. The authors created location profile for an identity, consisting of a pattern of zip codes of places a user geotagged. They measured similarity between two location profiles on the basis of cosine, Jaccard and other metrics. They developed timing profiles of two identities and similarity between two was measured on the number of timestamp matches. Authors also developed language profiles derived from textual content user created. The system was evaluated on Flickr, Twitter and Yelp, where location and timing profiles helped to narrow down the candidate set to search in for correct identity match while language models hit 94.7\% miss rate at 1\% false alarm rate~\cite{Goga}. This calls for a need to apply more sophisticated techniques and extract more intuitive content based features (e.g URLs, emoticons, spelling mistakes etc) for identity resolution. Some preliminary work was discussed in \cite{Gani} where authors presented the idea of applying authorship analysis techniques on the content created by user to find multiple identities of a user within a social network. Correa \textit{et al. } explored the URLs embedded in the public content posted by the user and observed that users often mention their identity on other networks via URLs posted to refer to their pictures, videos or other content. The authors introduced a system capable of finding identities of a user across networks by mentioning URLs in their content~\cite{Correa}.\\
\indent Apart from profile and content attributes, an attempt was made to link multiple identities by the third dimension i.e. network. Few de-anonymization techniques used network attributes to link one anonymized and one labeled user. Narayanan \textit{et al.} used a graph theoretic approach to de-anonymize Twitter users with the use of labelled Flickr network \cite{Narayanan}. The authors iteratively matched each node network to find to most similar node with the similar network and claimed 30.8\% accuracy. However the method needed a graph structure available on both networks, to deploy graph based algorithm proposed. In real world, network of a user is private giving an incomplete graph on any of the social network involved in identity search. In such cases, deanonymization algorithms have limited applicability. Labitzke \textit{et al.} also investigated the strengths of using network for identity linking. They suggested a different approach of matching mutual friends of two identities to be linked. The authors used string matching methods to link names of common friends of two identities. If there exist more than 3 mutual friends, the two identities were marked as linked (belonging to same person)~\cite{labitzke}. However, the approach had a gap of understanding that in real world, there could be multiple mutual friends between two profiles, or no mutual friends (in case when user used different social networks for different purpose). Further, it was assumed that complete network for both the identities was accessible and available, which may not be the case on each social network, thereby the applicability of the approach is limited. \\
\indent We apprehend that each dimension has been explored in the literature with few approaches benefiting from integration of more than one dimension. However, value of integration of all the three dimensions, to search for a user identity remains unexplored. To the best of our knowledge, we fill the gap by associating all three dimensions to work together, in order to search identity of a user on online social networks. We further limit our approach to access public attributes on any dimension, to avoid requirements for detailed information about an identity, for identity search and resolution. We are aware of few identity search systems e.g. Yasni, Pipl and PeekYou~\footnote{www.yasni.com, www.pipl.com, www.peekyou.com} which uses multiple approaches to search for a user on the Internet including social networks, however the approaches are not disclosed and candidate set returned for each user query are large, making the system unreliable and effort consuming. We leave the verification of this claim for our future work. Our work fills the gap of building a research based open integrated system to find for a user's multiple online identities.
\vspace{-3mm}
\section{Methodology} \label{Methodology}
In this section, we discuss the methodology proposed to search for a user identities across social networks. For a clear reference, the user whose identity is searched is termed as ``\textit{Nemo}". We explain a set of algorithms which exploit publicly available information of nemo on Twitter, to search for her identity on Facebook. The algorithms are -- { \textbf {Profile Search}}, {\textbf{Content Search}, \textbf{Self-mention Search}} and {\textbf{Network Search}}. The algorithms access only publicly available data about any user, as compared to other algorithms proposed in literature which were allowed to access detailed information about a user as discussed in Section~\ref{Related}. We now discuss each of the algorithms in detail.
\vspace{-3mm}
\subsection{Profile Search}\label{Profile}
\begin{figure}[ht]
  \centering
  \includegraphics[page=1,scale=0.25]{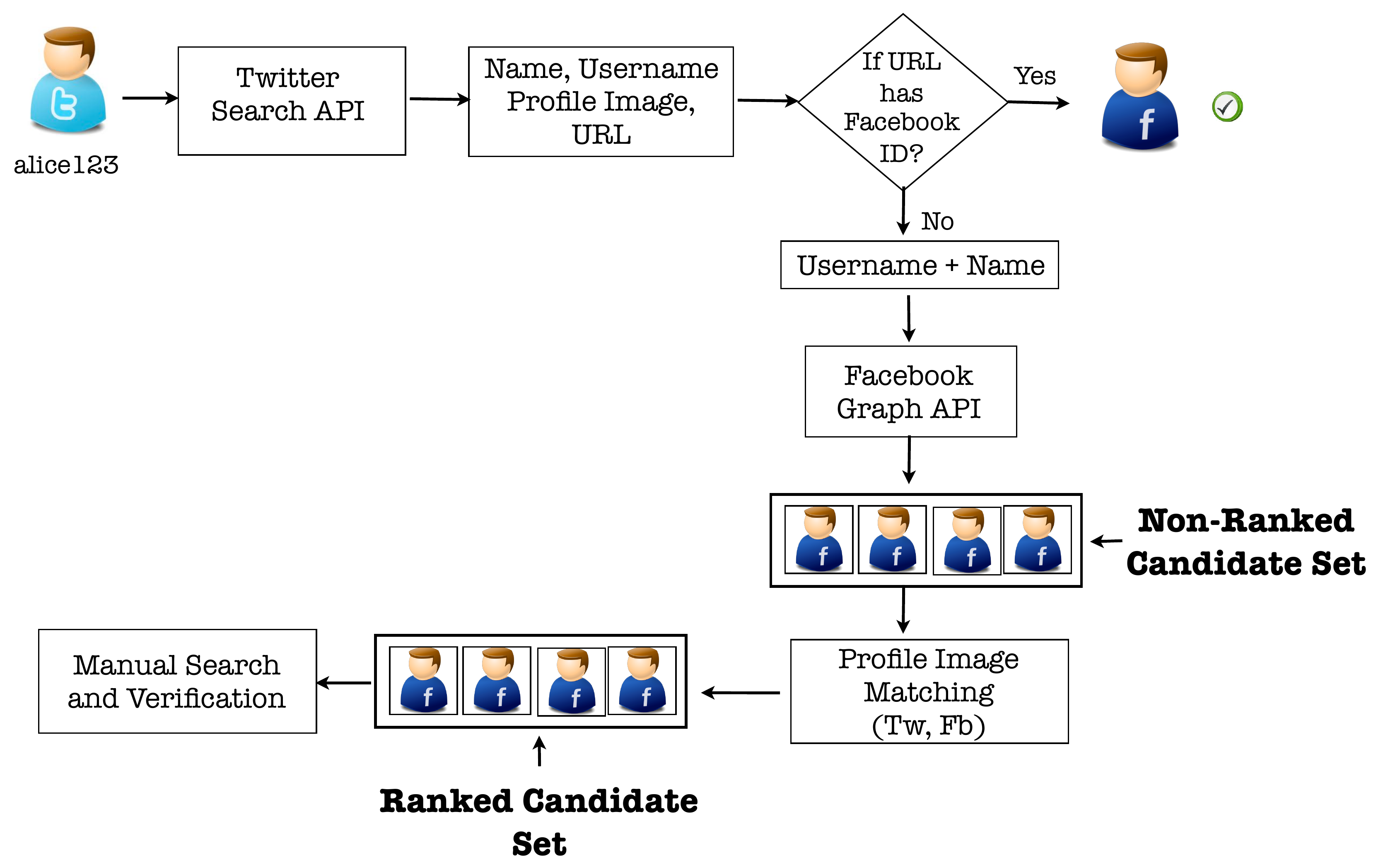}
  \caption{Profile Search Algorithm. In this method, we use profile information of a user to locate her identity on Facebook. }
  \label{fig:profile}
\end{figure}

\indent An identity of a user on a social network includes a set of profile attributes, which gives basic information about the user such as username, name, location, gender, description etc. If the user does not demonstrate any active obfuscation and does not create altogether a different identity, it is likely that she re-uses certain profile attributes' value, on the social networks she joins. If the user demonstrates such behavior, profile attributes can be used to find her identity on other social networks. Profile Search method explores ``Profile" dimension of a user's identity and exploits profile attributes as linking attributes. To make comparisons between any two identities using linking attributes, it is essential to have same set of attributes publicly available for both identities. Twitter has a limited set of attributes however publicly available~\footnote{Accessible to any user on the Internet.} while Facebook has larger set of attributes, however private. We consider only those profile attributes which are publicly available on both networks -- username, name, profile image and URL. Using the value of these attributes for nemo on Twitter, we search Facebook. We add location as another attribute available on Twitter to refine the search on Facebook. The search produce a list of candidate identities with same attribute values as of nemo on Twitter. The flow of the Profile Search algorithm is illustrated in Figure~\ref{fig:profile}. \\
\indent Firstly, we use nemo's username on Twitter, and query Twitter API to extract her name, username, location, profile image and URL. We use URL attribute first to observe if nemo herself has given her Facebook identity. We term this behavior of mentioning one's Facebook network identity (or any other network identity) on Twitter explicitly, as ``\textit{Self-Identification}". We observed two varieties of self-identification behavior -- one in which a user directly gives her Facebook identity on her URL attribute and other in which a user indirectly gives her Facebook identity via referring to a webpage on her URL attribute, that contains her Facebook identity. A user referring to her blog on Twitter URL with her blog having her Facebook identity is an example of indirect self-identification. If nemo has not identified herself via URL, we use her username, name and location attribute to query Facebook Graph API to find identities with same or similar username / name having the same or similar location. Facebook Graph API returns a set of searchable~\footnote{Users who allow to be searched within Facebook and do not have this feature turned off in privacy settings.} identities (users, pages and communities) who either have same name as the ``queried" name or a part of ``queried name" in their name and share ``queried" location.~\footnote{``Queried" name is nemo's name on Twitter.} We also search for a candidate identity on Facebook who has the same username as nemo's Twitter username. The reason for this addition is that previous research shows that around 66\% of users use same / similar usernames across social networks~\cite{Correa}. Therefore, there is a possibility of nemo using the same username on Facebook as on Twitter. We aggregate nemo's candidate identities on Facebook as returned by Facebook Graph API and term the set as ``Non-ranked" set.  To further ease the search of nemo's identity in the aggregated set, we rank the candidate identities on the basis of nemo's Twitter profile image and the candidate identity's profile image similarity, using RGB histogram matching. It is highly probable that nemo might use same profile picture on more than one networks. On the ranked candidate set, a manual verification is done to accurately locate nemo's Facebook identity. 
\vspace{-3mm}

%We then feed the result set in our ranking engine, which returns the ranked results on the basis of ranking parameters. Ranking engine rank results on the basis of similarity between Twitter profile image and other network profile image and on the basis of bigram matching similarity of Twitter username and network username.
\subsection{Content Search}
\begin{figure}[!t]
  \centering
  \includegraphics[ page=2,scale=0.23]{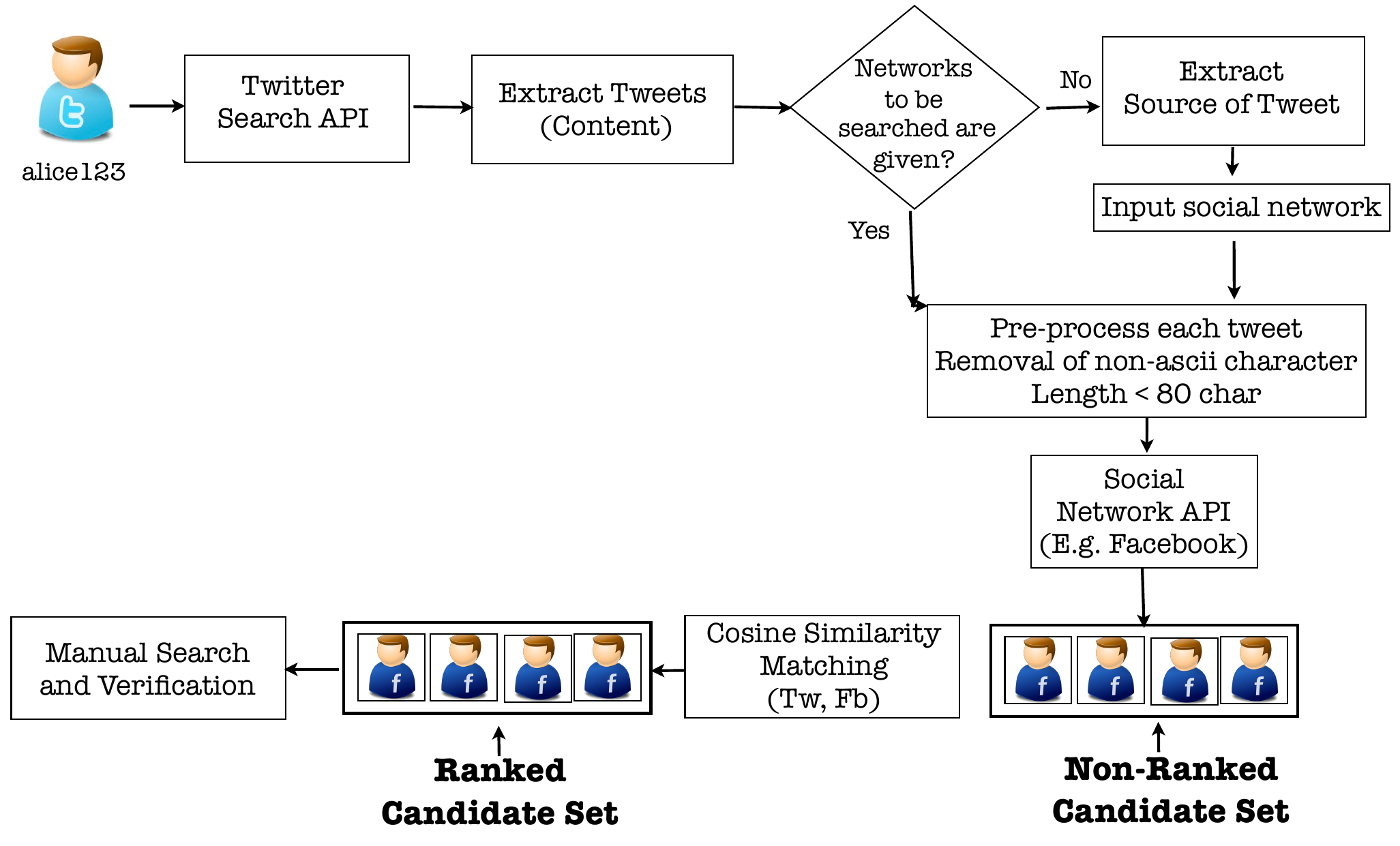}
  \caption{Content Search Algorithm. In this method, we use content of a user to locate her identity on Facebook. We search for Facebook users who posted / shared similar content as tweeted by nemo.}
  \vspace{-3mm}
  \label{fig:content}
\end{figure}

\indent An identity of a user on a social network includes the content that she creates or is shared with her. Owing to the popularity of social aggregation sites and ways to link multiple networks together, a user pushes the same content on multiple networks simultaneously. For example, Twitter provides a functionality to connect Twitter and Facebook identity to post user's tweets on her Facebook timeline, Twitterfeed~\footnote{http://www.twitterfeed.com} allows a user to connect Twitter, Facebook, LinkedIn to push feeds in three social networks simultaneously. Because of such services, it is likely that a user generates same content on multiple social networks. Such a user behavior can be exposed by Twitter API which provides the ``source" of a tweet i.e. from where the tweet is posted e.g. Facebook, Twitterfeed etc.  Source can be exploited to reduce the search space for a user's online identities, if an analyst intend to save her efforts by searching for a user in only social networks where she has hints of her existence. Content Search method uses content as a linking attribute for users who use the mentioned services. In this paper, we do not use source of the tweets since we limit our focus to search for nemo's identity only on Facebook and with the help of ground truth we know the nemo has a Facebook identity. However, we plan to use this information in our future work. \\
\indent Figure~\ref{fig:content} explains the flow of content search algorithm. We extract most recent 100 (or less)~\footnote{We limit to process most recent 100 tweets to avoid long execution time.} posts by nemo, and process each of the posts to limit the length to 75 characters and to remove non-ascii characters. We query Facebook Graph API with the processed post to search for the users who posted same or similar content on Facebook. Facebook Graph API returns a candidate set of Facebook identities of users who posted similar content as queried content, which is then fed into our ranking engine. Ranking rank the candidate identities on the basis of cosine similarity score between the nemo's tweet terms and candidate identity's post terms. A manual verification is further needed to hover on each candidate profile, and search for nemo's correct Facebook identity.
\vspace{-3mm}
\subsection{Self-mention Search}
\begin{figure}[ht]
  \centering
  \includegraphics[page=4,scale=0.23]{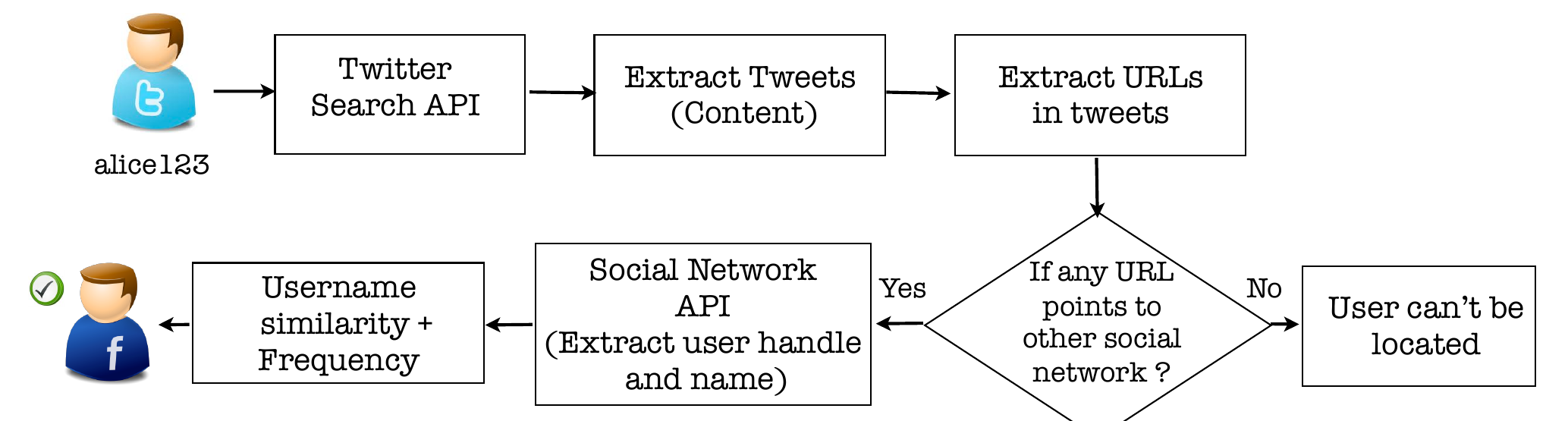}
  \caption{Self-mention Search Algorithm. In this method, we use content information of a user to observe if a user herself has posted a link to her Facebook post / identity.}
  \label{fig:self-mention}\vspace{-3mm}
\end{figure}

\indent This method exploits a user's tendency to cross-pollinate information on Online Social Media~\cite{Jain} and was introduced by Correa \textit{et al.} \cite{Correa}. The method explores content attributes of nemo and assumes that if nemo has accounts on two or more networks, she might cross refer to her other account, in few of her tweets. For example, nemo might post a tweet with a URL referring to an album on Flickr, indirectly revealing her Flickr identity. We term this behavior of posting URLs indirectly but consciously, pointing to user's other network identity as ``\emph{Self-mention}". Self-mention behavior allow identity leaks via content created in the form of URLs by the user.  This method exploits self-mention behavior to search for a user identities across networks.\\
\indent Figure~\ref{fig:self-mention} illustrates the process. We query Twitter Search API to extract 100 (or less) recent tweets by nemo and filter out the tweets with URLs and then further process each URL to verify if it refers to Facebook. We create a set of all the Facebook URLs posted by nemo, query Facebook Graph API to process each URL and extract name and username of the candidate user (if the URL refers to a user's identity and not apps). We then calculate candidate identity's username similarity with nemo's Twitter username to rank the candidate set. A manual verification of top rank candidate identity and most frequently referred candidate identity decides nemo's Facebook identity.
\vspace{-3mm}
\subsection{Network Search}
\begin{figure}[htbp] %  figure placement: here, top, bottom, or page
   \centering   
    \vspace{-3mm}
     \includegraphics[page=3,scale=0.23]{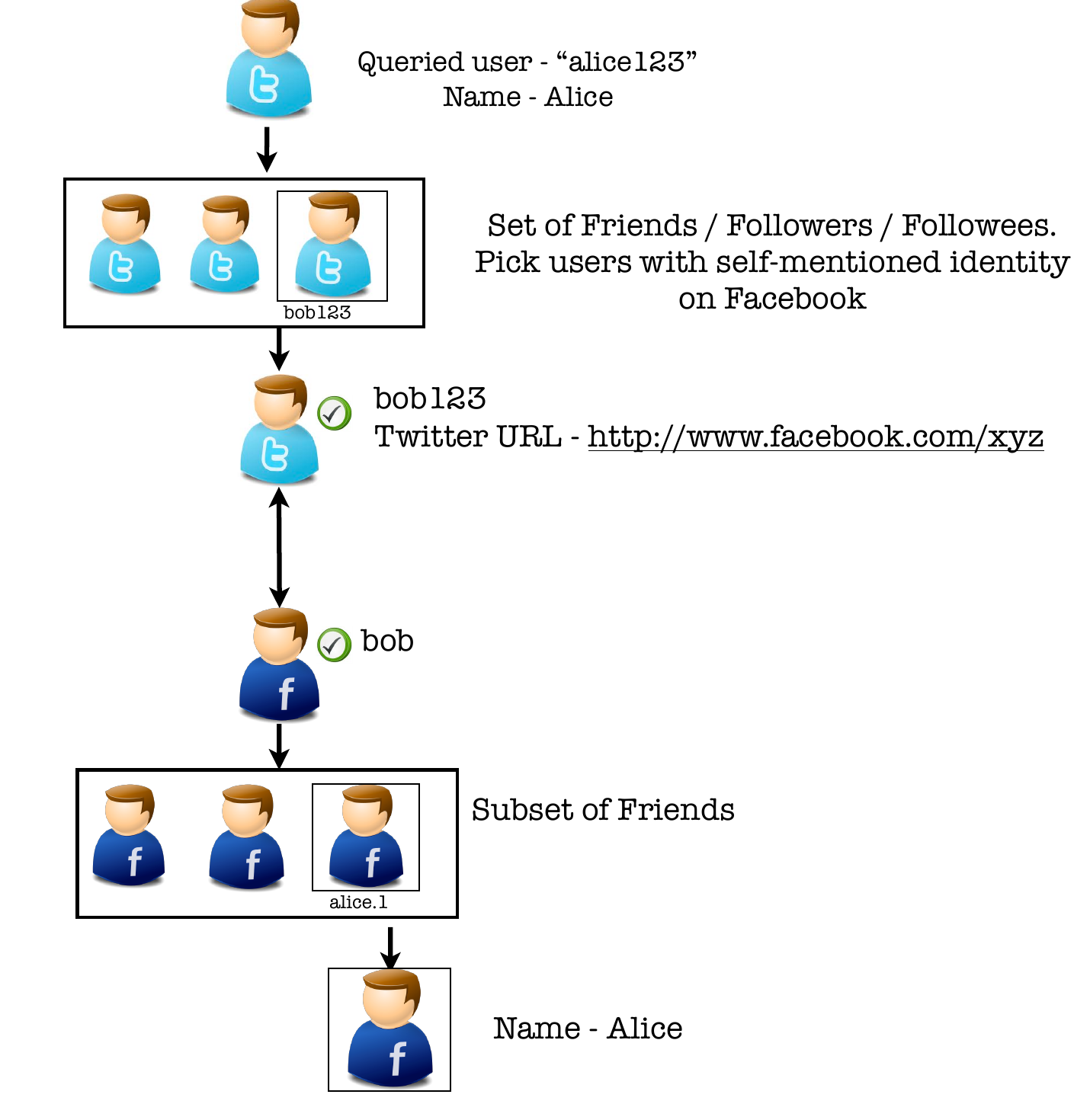}
     \vspace{-3mm}
   \caption{Network Search Algorithm. In this method, we use nemo's Twitter network to locate her identity on Facebook. \vspace{-3mm}}
   \label{fig:network}
\end{figure}

\indent Network is an important dimension of a user's identity on a social network. It is a shared identity of a user build with the involvement of other users~\cite{Rowe10}. If other users leak their identity on any other social network, it is likely that the user's identity also gets leaked. Network Search algorithm explores the possibility of an user's identity leak via her network attribute.   \\
\indent We search for nemo's identity on Facebook using her Twitter network. We consider her three Twitter networks -- follower network, followee network and friend network. Follower network is composed of users who follow nemo and has her least control in creation of the network. Followee network includes users whom nemo follows and has her huge control in creation of network. Friends network intersects both followee and follower network and includes users who follow nemo as well as are followed by her. Friend network on Twitter can be considered as bilateral network, similar to friends network on Facebook, with both users involved actively in the decision to build the relationship. For each of the three networks, we filter out those who have self - identified themselves on Facebook via their Twitter URL. In this way, we map nemo's network from Twitter to Facebook. We now assume that nemo connects to a same subset of users on both social networks. We search for nemo's name in public Facebook friend-list of mapped Facebook identities of nemo's Twitter network users. If a Facebook identity with same / similar name as nemo is friends with more than one nemo's follower / followee / friend identity on Facebook, the Facebook identity is claimed to be nemo's Facebook identity (see Figure~\ref{fig:network}). In this way, we try to reverse map nemo's identity from one social network to another via mapping her network on two social networks.  Note that the method is applicable, even when the complete network   of nemo's network or the network structure of nemo's network on Facebook is partially available.\\
\indent In a nutshell, we experiment with all the three major dimensions of a user's identity on a social network. We observe that some users consciously give their Facebook identity by self-identification, and self-mention while other users are uninformed with no intensions of giving her Facebook identity e.g. identity leak via name, location, content and network. We now evaluate each of the algorithms discussed to understand its effectiveness in finding a user online identities.
\vspace{-2mm}
\section{Evaluation} \label{Evaluation}
To evaluate the algorithms, we borrowed a ground truth dataset from \cite{Malhotra} collected from Social Graph API. The dataset consisted of 505,466 users who themselves mentioned their identity on multiple social networks including Twitter and Facebook. To avoid bias in the evaluation either towards a dataset of users with multiple identity leaks or a dataset of users with few identity leaks, we randomly selected 543~\footnote{We are continuously running the experiments. We report the till date dataset and evaluation results.} users from the bigger dataset (505,466) and evaluated the algorithms. We measured the efficiency of the  algorithms on three parameters --  \textit{Accuracy}, \emph{Candidate set size}, and \emph{Search time}. We define each of the parameters below -
\vspace{-2mm}
\begin{itemize}
\item 
\vspace{-2mm}
\textbf{Accuracy} - The metric is defined as ratio of users for whom correct Facebook identity is returned and users for whom Facebook identity is searched. Higher the accuracy, better is the algorithm.
\item
\vspace{-2mm}
\textbf{Candidate set size} - The parameter measures the number of candidate identities returned for a searched Twitter user. Smaller the candidate set size, easier is for an analyst to search in for correct Facebook identity of the user. 
\item
\vspace{-2mm}
\textbf{Search time} - It is defined as time taken to return the candidate set for a searched Twitter user. Lesser the time, faster is the algorithm.
\vspace{-3mm}
\end{itemize}
\indent We evaluated the algorithms on a Ubuntu server with six quad core processors each of 1.87GHz speed, 16Gb RAM, 8Gb cache size.
\vspace{-3mm}
%The success of the algorithms largely depends on the nature of identity leak (profile, content, network) for a user. If the user searched has identity leaks, the algorithms can exploit the leak and find the correct Facebook identity. 
\subsection{Profile Search}
\subsubsection{Accuracy}
With Profile Search, we could locate Facebook identities for 205 users (37.7\%) within the candidate set returned, via URL mentioned by user herself, by search with the same username on Facebook or search via user's name, username and location specified on Twitter. We further quantified the number of users located correctly on Facebook 
by each of the three methods and their combinations. Table \ref{tab:booktabs1} gives the numbers. We observed that 25.2\% of the 543 users had consciously mentioned their Facebook identity by mentioning it on Twitter's URL attribute (directly or indirectly as discussed in Section~\ref{Profile}). However, for 27.4\% of users who did not mention their Facebook identity as URL, the method identified their Facebook account by either searching for same username or searching user's name, username and location. An uninformed and unintended identity exposure via their profile attributes gave away a user's identity on other social network.
\begin{table}[htbp]
   \centering
   \begin{tabular}{|l|p{1.35cm}|p{1.35cm}|} % Column formatting, @{} suppresses leading/trailing space
     \hline
      \textbf{Search Method}   & \textbf{Accuracy (out of 543)} & \textbf{Accuracy in percentage} \\ \hline
         URL Search (URL)     & 137 & \textbf{25.2\%} \\ \hline
        Same username Search (SU)   & 82  & 15.1\% \\ \hline
       Name Location Search  (NL)    & 144 & 26.5\% \\ \hline
        URL+ SU     & 175 & 32.2\% \\ \hline
        URL + NL   & 200  & 36.8\% \\ \hline
       	SU + NL       & 149  & \textbf{27.4\%} \\ \hline
        URL +  SU + NL     & 205 & \textbf{37.7\%} \\ \hline
                 \end{tabular} 
   \caption{Profile Search  Evaluation. We observed that 25.2\% of the users had mentioned their Facebook identity in the URL on Twitter, however even without that information, we were able to find identity on Facebook for 27.4\% of the users. }
   \label{tab:booktabs1}
\end{table}
\vspace{-4mm}
\subsubsection{Candidate set size}
We analyzed the number of candidate Facebook identities returned by Profile Search. Figure~\ref{fig:cand} shows the distribution of Facebook candidate identities set size for the users whose correct identity was located within the candidate set (205 users). We observed that for most of the users, candidate set size varied from 0 to 20~\footnote{We queried Graph API to return a maximum of 60 candidate identities for a user.} We then analyzed the affect of ranking candidate identities on the basis of profile image matching, in order to reduce the candidate set size to search in for user's Facebook identity. We investigated the rank at which correct identity of the user is returned, both for the ranked candidate set and non-ranked candidate set. We analyzed 205 users and observed that  correct identity of a user could be located at lower rank with ranking as compared to without ranking. We observed that for 98\% of the users, correct identity on Facebook was located within top-5 candidates of non-ranked candidate set with 2\% users located in top-25 candidates while for 99.51\% of the users, correct identity was located within top-10 ranks of ranked candidate set. Hence, an analyst had to scan only top-10 results, irrespective of the candidate set size returned by method, for 99.51\% of the users for whom Facebook identity was recognized correctly.
\vspace{-3mm}
\subsubsection{Search time}
We analyzed the time taken to search for the user via profile attributes and for majority of the users, search time is less than a minute. Time taken to rank the candidate set was negligible. However, we restricted the Graph API to return maximum of 60 candidate identities for a user. If we had extended the limit, we believe that search time could have increased. We left the verification of the claim to our future work.\\
% Add what if you call to return 40 results. Does the accuracy increases? For now, the limit is 20
% \begin{figure}[!ht] %  figure placement: here, top, bottom, or page
%        \subfigure{
%         \includegraphics[scale=0.21]{norank.pdf}
%      
%         \includegraphics[scale=0.21]{rank.pdf}}
%	
%       %  \includegraphics[scale=0.2]{rank.pdf} 
%         %\label{fig:rank}}
%         
%    \caption{Evaluation of profile image based ranking. We observed that with ranking, few users moved nearer to top results helping an analyst in the identification of the user.}
%    \label{fig:ranking}
% \end{figure}
\indent Therefore, for a Twitter user searched, Profile Search algorithm returned a larger set of candidate Facebook identities within a minute.  Correct Facebook identity was identified for 37.3\% of user and located within top-10 candidate identities by manual verification.\vspace{-3mm}
\subsection{Content Search}
\subsubsection{Accuracy}
We evaluated Content search algorithm with 543 users and received a set of candidate Facebook identities for 124 users only (22.8\%). When searched for the correct identity out of the candidate identities for a user, only three users' Facebook identities were correctly identified, reporting an accuracy of 0.5\%. Accuracy of the algorithm turned out to be low (0.5\%) 
owing to multiple reasons. Firstly, only a limited number of users (22.8\%) received a candidate set of Facebook identities. The reasons could be -- users did not post same / similar content across networks or the users did post same / similar content across networks, however their content on Facebook was not publicly available and searchable. Secondly, candidate Facebook identities were returned because of the popular content tweeted by nemo and posted by hundreds of users on Facebook (e.g. quotes, sayings). It was highly probable that nemo's correct Facebook identity was not returned in the candidate set.~\footnote{It depends on Facebook API recall for a search query.} However, there were few Twitter users for whom the corresponding Facebook identity posted same content with her friends only but her friends re-posted the same content and shared with public, making the content searchable and the correct Facebook identity could then be identified. We sense an information privacy leak for the user via her friends re-sharing activity. \vspace{-3mm}
\subsubsection{Candidate set size}
We analyzed the candidate set size returned for the users for whom correct Facebook identity is returned. We observed that the candidate set size ranged from 8 - 30 (see Figure~\ref{fig:content_cand}). The candidate set size could have been larger if more than 100 posts of a user were queried to find the Facebook identity. \vspace{-3mm}
\subsubsection{Search time}
Content Search method returned the candidate set within 20 secs for the three users correctly identified on Facebook, faster than the Profile Search method (see Table~\ref{tab:booktabs6}). Lower search time could be attributed to few tweets available to search on Facebook for all three users. \\
 \begin{table}[htbp]
   \centering
   %\topcaption{Table captions are better up top} % requires the topcapt package
   \begin{tabular}{|l|p{2cm}|p{2cm}|} % Column formatting, @{} suppresses leading/trailing space
     \hline
      \textbf{User}   & \textbf{Profile Search} & \textbf{Content Search} \\ \hline
          User 1& 46.0s & 15.6s\\ \hline
           User 2 &  40.2s & 12.6s \\ \hline
            User 3 &  3.8s & 15.4s \\ \hline
          \end{tabular}
   \caption{Comparison of Profile Search and Content Search on the basis of Search time.}
   \label{tab:booktabs6}
\end{table}
\indent We believed that success of the algorithm was dependent on the availability of user posts on both networks. To verify the belief, we experimented with another randomly selected sample from the bigger dataset consisting of 356 users. We repeated our Content Search evaluation and located 12 users' Facebook identity correctly (accuracy of 3.4\%). Therefore, accuracy of the method was highly dependent on the presence of users in the evaluation dataset who leak their identity via content as well as for whom the content is publicly available on both networks.\vspace{-3mm}
\subsection{Self-mention Search}
\subsubsection{Accuracy}
We experimented with 543 Twitter users, and searched in most recent 100 tweets of a user. Only 31 Twitter users' Facebook identities were correctly identified (accuracy of 5.7\%) via URLs posted by them. On further manual investigation, we observed that most of the users posted URLs pointing to either Facebook profile page or her Facebook post / video / picture, thereby revealing their Facebook identity on Twitter indirectly, consciously. \vspace{-3mm}
\subsubsection{Candidate set size}
\begin{figure*}[!t] %  figure placement: here, top, bottom, or page
    \mbox{\subfigure[Profile Search]{
        \includegraphics[scale=0.2]{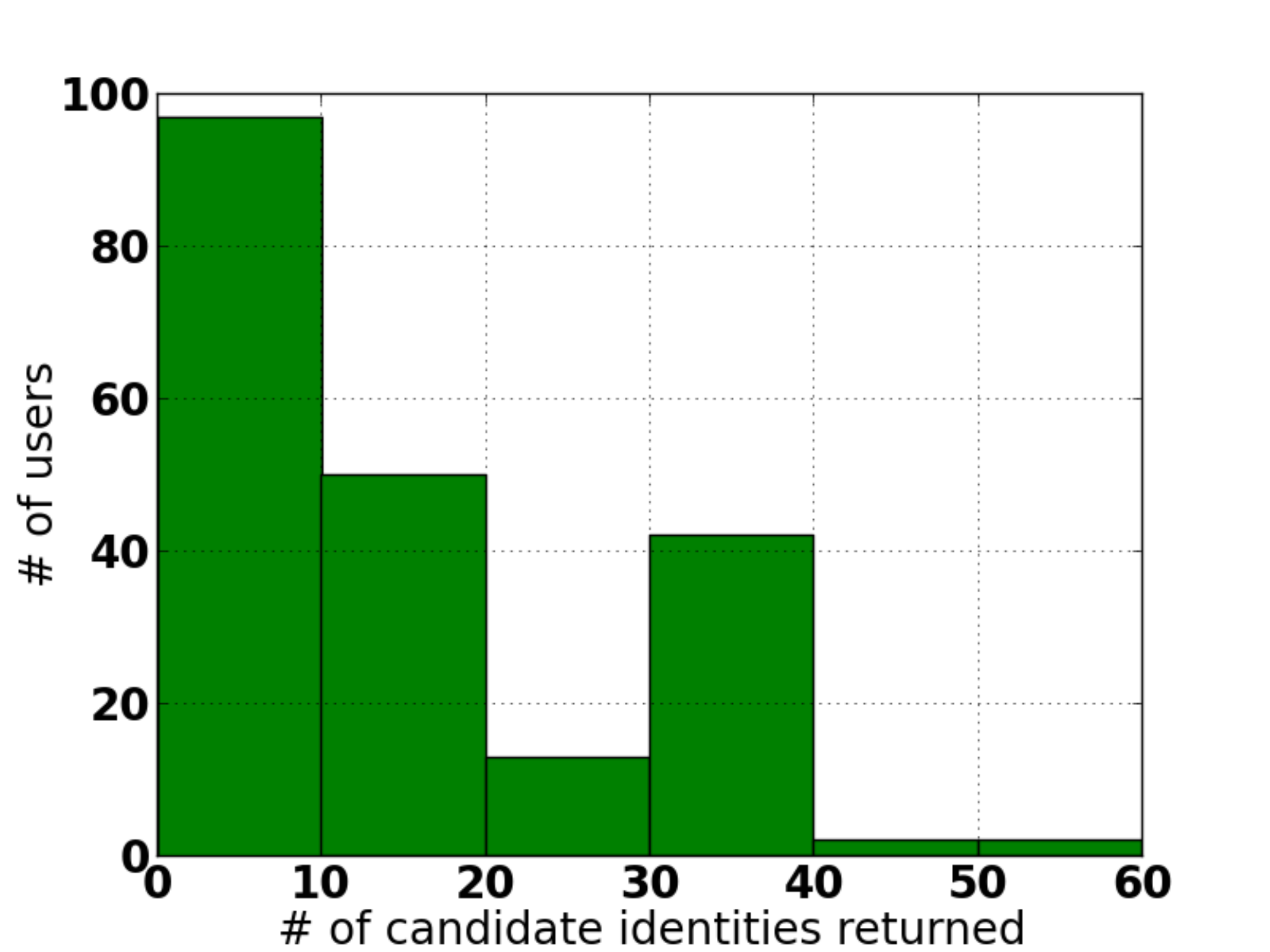}
    \label{fig:cand}}
    \quad  
  
     \subfigure[Content Search]{
        \includegraphics[scale=0.2]{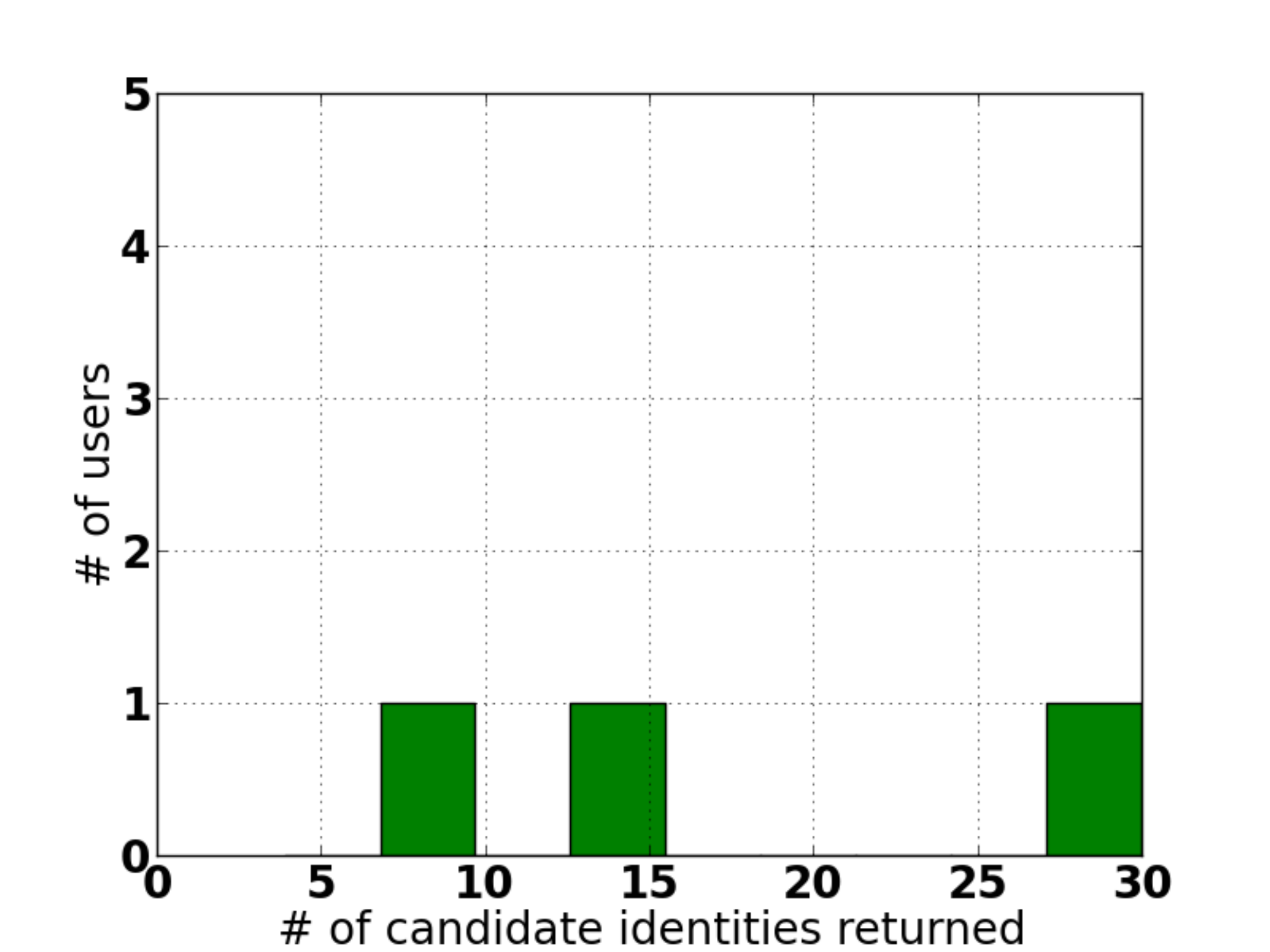}
    \label{fig:content_cand}
    }
    \quad
    \subfigure[Self-mention Search]{
        \includegraphics[scale=0.2]{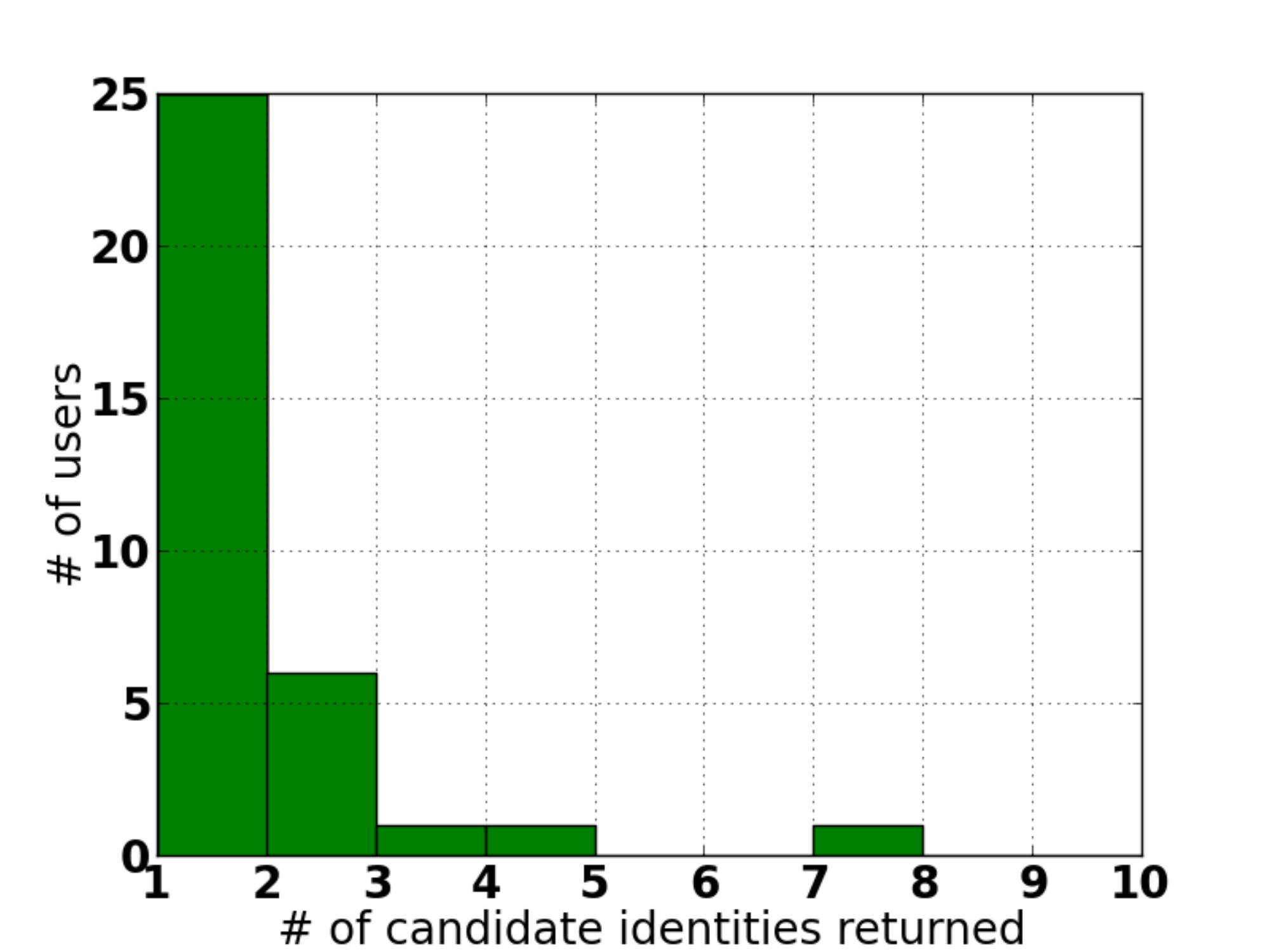}
    \label{fig:w_cand}
    }
        \quad
    \subfigure[Finding Nemo]{
        \includegraphics[scale=0.2]{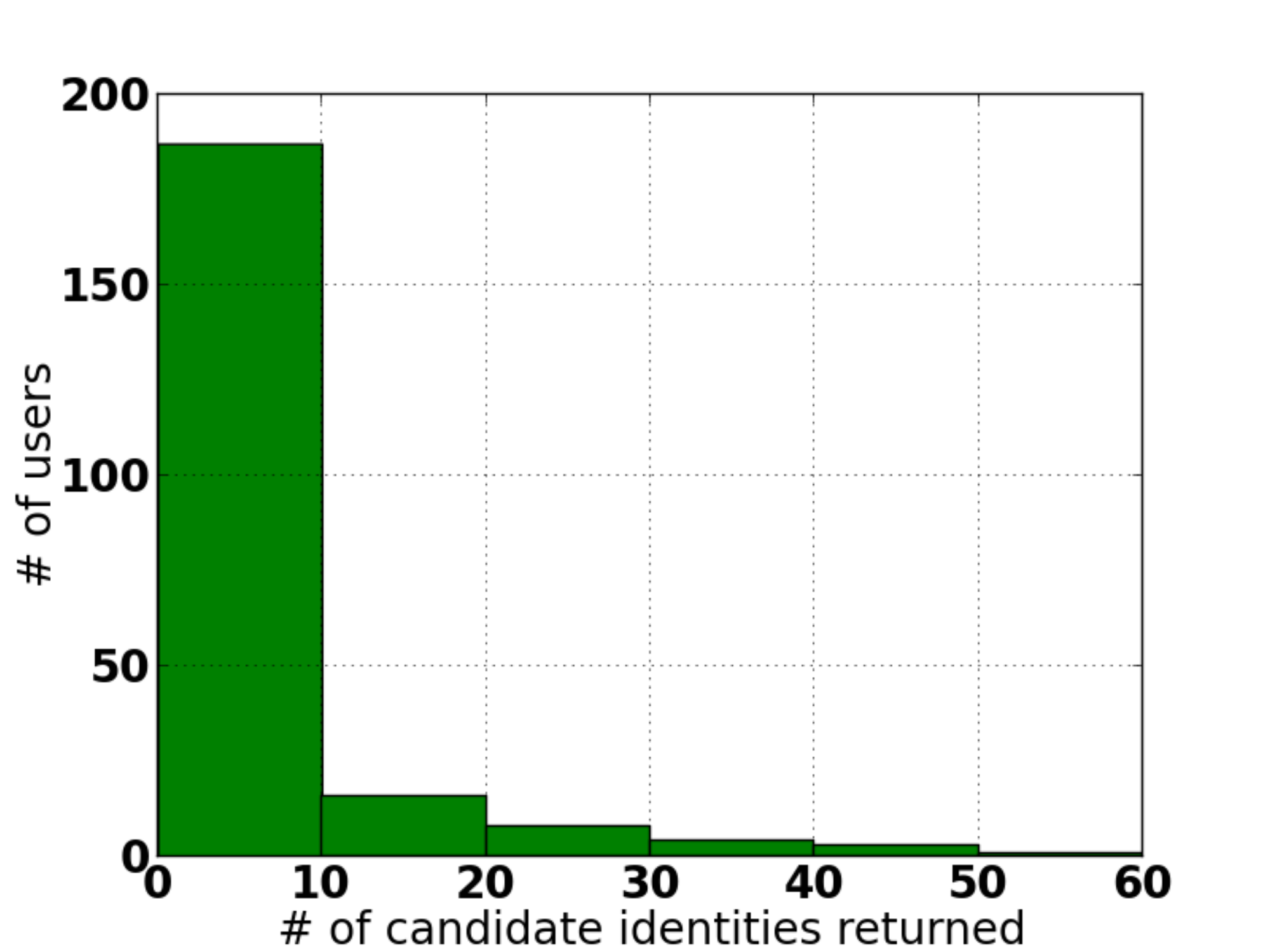}
    \label{fig:n_cand}
    }
    }  \vspace{-2mm}
    \caption{Evaluation of Profile Search, Content Search, Self-mention Search method and Finding Nemo on the basis of candidate set size returned. }
         \vspace{-2mm}
    \label{fig:candidate}
 \end{figure*}

Self mention Search largely reduced the candidate set size as shown in Figure~\ref{fig:w_cand}. For majority of the correctly identified users, the correct identity was located within top-3 results. Therefore, an analyst had to search in smaller candidate set, thereby proving the effectiveness of the algorithm. \vspace{-1mm}
\subsubsection{Search time}
\begin{figure*}[!tp] %  figure placement: here, top, bottom, or page 
\mbox{
       \subfigure[Profile Search]{  
         \includegraphics[scale=0.2]{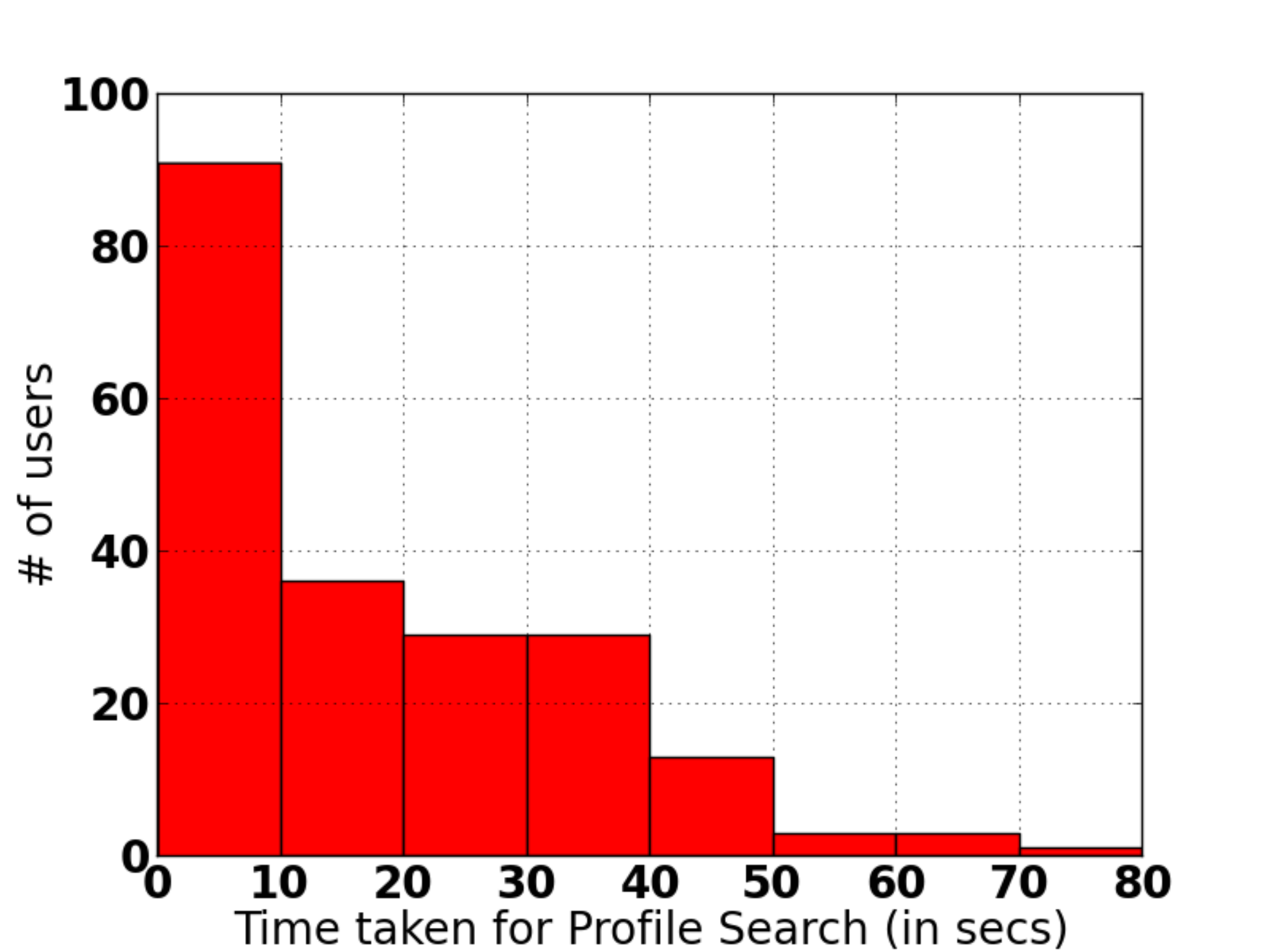}
         \label{fig:time}
    }
\quad
   \subfigure[Content Search]{
      
         \includegraphics[scale=0.2]{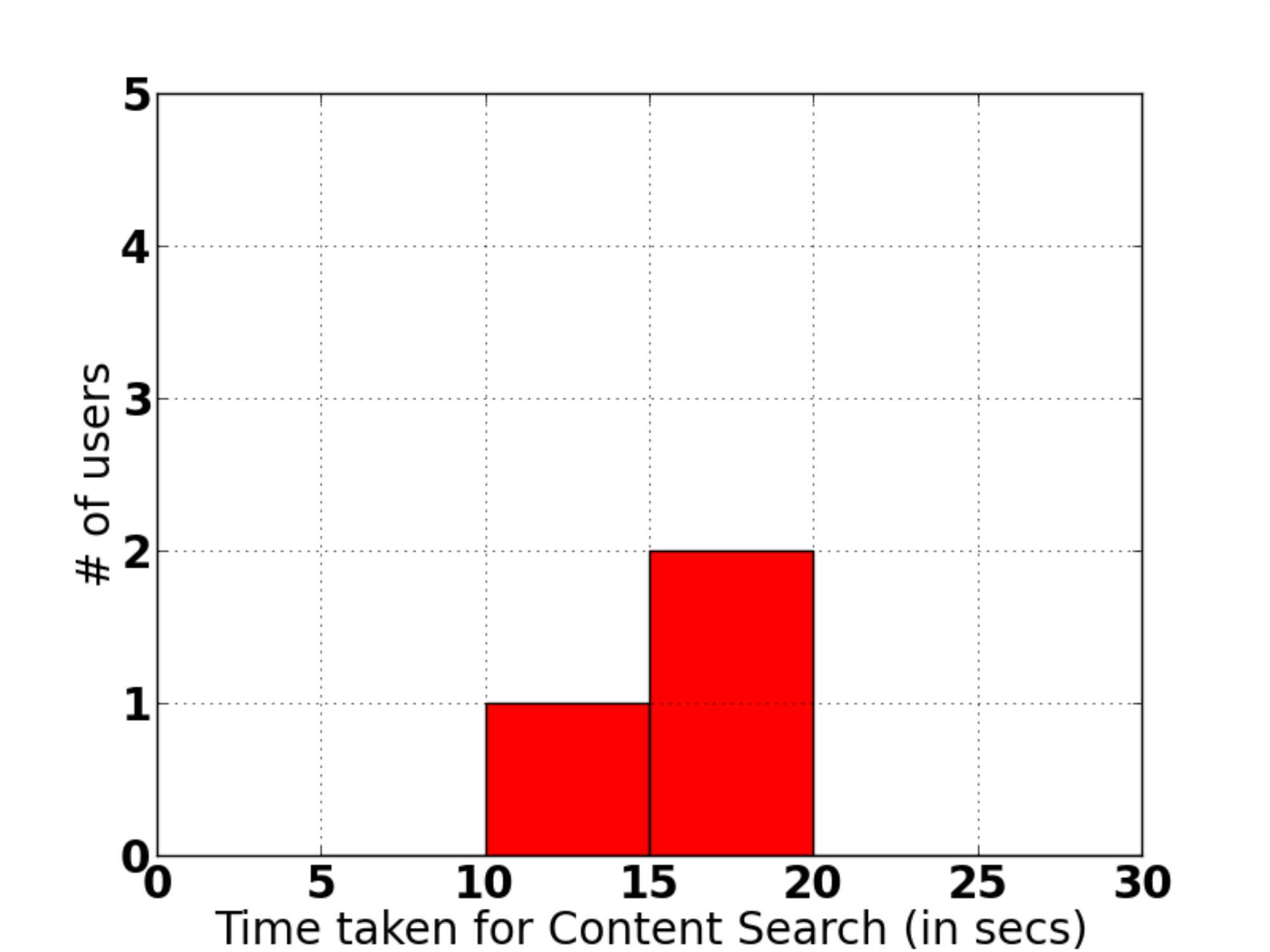}
         \label{fig:c_time}
    }
\quad
        \subfigure[Self-mention Search]{
      
         \includegraphics[scale=0.2]{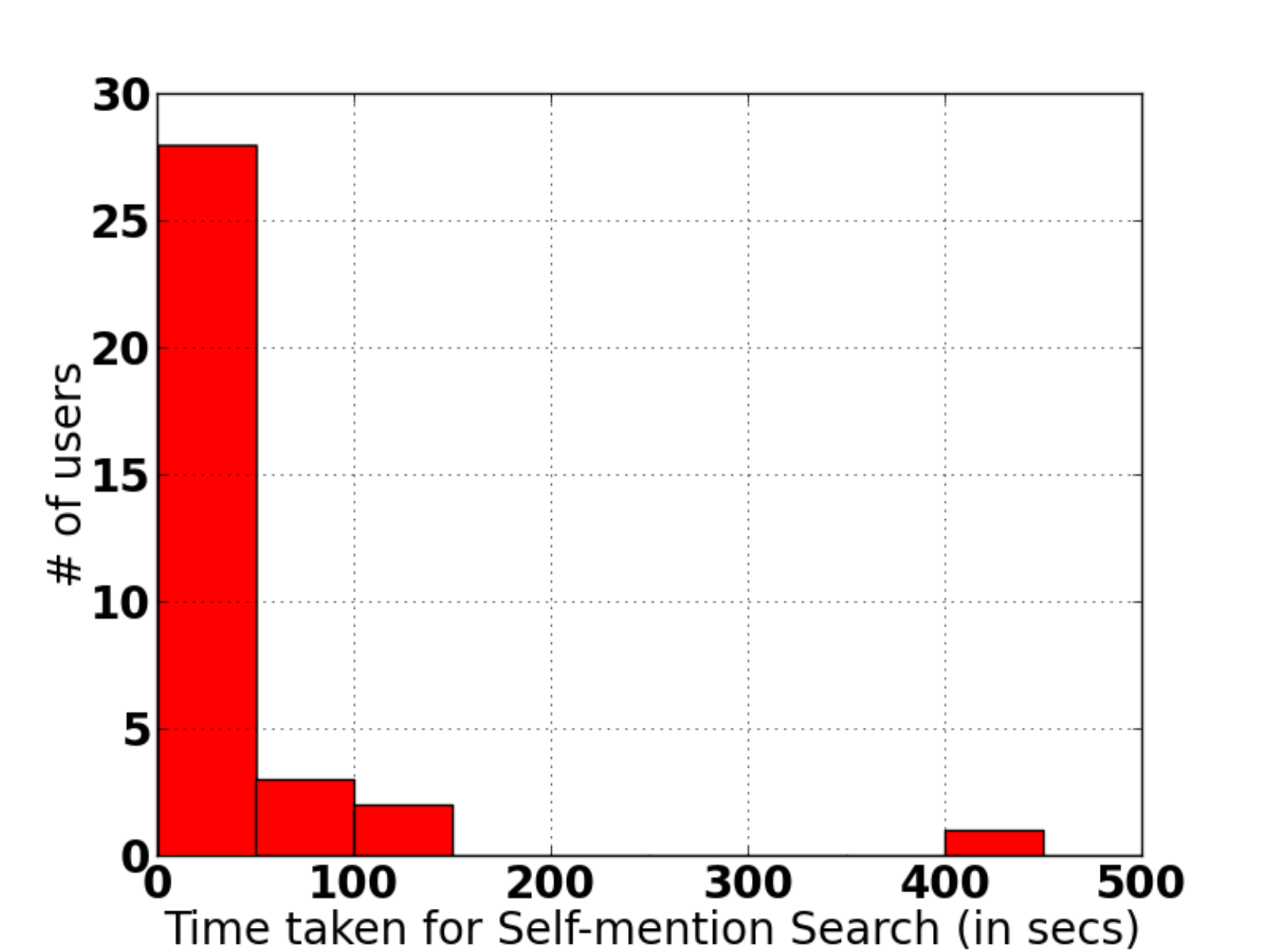}
         \label{fig:w_time}
    }
    \quad
        \subfigure[Finding Nemo]{
      
         \includegraphics[scale=0.2]{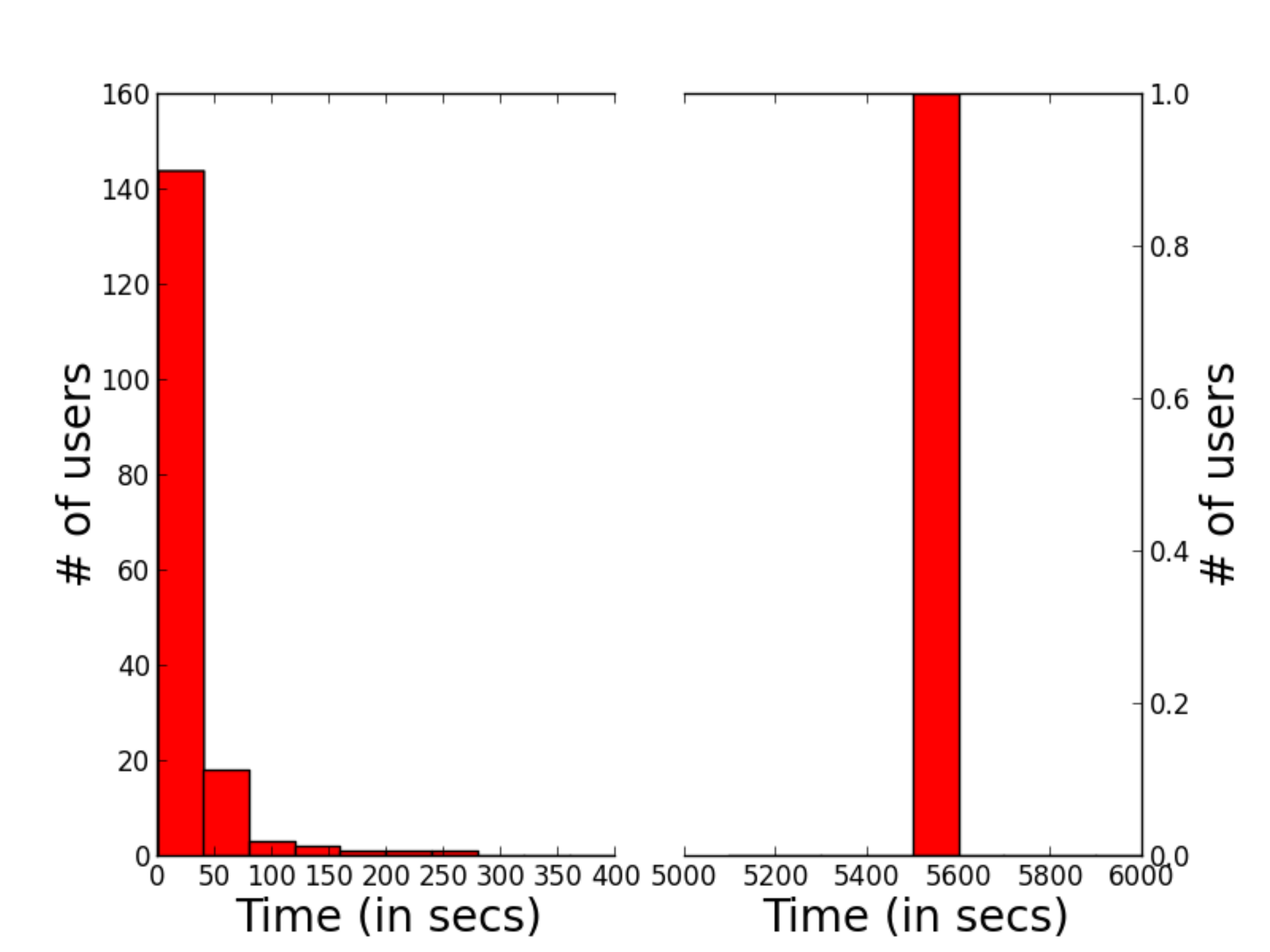}
         \label{fig:nemo_time}
    }
  }
  \vspace{-2mm}
    \caption{Evaluation of Profile Search, Content Search, Self-mention Search method and Finding Nemo on time taken to generate candidate set. }
      \vspace{-3mm}
    \label{fig:timetaken}
 \end{figure*}

We analyzed the time taken to return the candidate set and observed that self-mention search took more time as compared to Profile and Content Search (ranging from 1 - 150s) for majority of the users (see Figure~\ref{fig:w_time}). The reason for the higher search time was the presence of URL in majority of user's tweets. Each URL was processed to extract Facebook identity from a given URL. URL resolution process was time-expensive and repeating the process for each URL added to higher search time for Self-mention algorithm. \\
\indent Therefore, Self-mention search revealed Facebook identity for 5.7\% of the users within top-3 candidate identities at the cost of more search time.
 \vspace{-3mm}
\subsection{Network Search}
\subsubsection{Accuracy}
We experimented with same set of 543 Twitter users to evaluate Network Search algorithm with three networks (followers, followee and friends) to understand which network has the highest capability to expose a user's Facebook identity. We observed that network search exposed only one user's Facebook identity via her follower network. The reason for low accuracy was that even though people in the user's Twitter network mentioned their Facebook identity on Twitter, their friend-list on Facebook was not public and therefore, only few users could be located. No users were identified by followee network since the network majorly included celebrities, brands and events which had corresponding page identities on Facebook with no friends attribute. We tried to extract users who liked and subscribed to pages and community, but the information was not publicly available. No users' were identified by friends network owing to the restricted access to their Facebook friend-list. \\
%Table shows a comparison of accuracy achieved by Network Search algorithm with each of the three networks.
%
% \begin{table}[htbp]
%   \centering
%   %\topcaption{Table captions are better up top} % requires the topcapt package
%   \begin{tabular}{|p{2cm}|p{1.5cm}|p{1.5cm}|l|} % Column formatting, @{} suppresses leading/trailing space
%     \hline
%        \textbf{Network}  & \textbf{Followers} & \textbf{Followees} &  \textbf{Friends}    \\ \hline
%           \# of users identified & 6 (run) & 0 & 0 \\ \hline
%          \end{tabular}
%   \caption{Comparison of three networks for Network Search Algorithm. We observed that followers network was more effective in locating a user's Facebook identity as compared to other networks.}
%   \label{tab:booktabs4}
%\end{table}
\indent To verify the failure of the algorithm, we evaluated Network search algorithm on a different dataset. The dataset consisted of 10 users from India whose correct Twitter and Facebook identity were recorded manually with the help of authors' offline information about users. We executed Network Search on this dataset and found Facebook identity for seven of them. Varying accuracies on different datasets disprove the failure of the algorithm. On this basis, we accept that network search algorithm fails when the required information to find a user's Facebook identity is unavailable and restricted, however we emphasize on the non-zero accuracy of the method, verifying that the method can be exploited to leak a user's identity.
\vspace{-3mm}
\subsubsection{Candidate set size}
For the one user correctly identified on Facebook, the method returned only one candidate identity, which is the correct Facebook identity itself. The method reduced the candidate set size largely and increased the precision of finding the correct Facebook identity in top-1 candidates. However, we do not generalize the observation. We think that since the method involved specific user attributes (network) which is likely not same for a large set of users, the method has the potential to return correct Facebook identity as the candidate set. \vspace{-3mm}
\subsubsection{Search time}
Network Search algorithm returned a precise candidate set at the expense of huge search time. The method took more than 2 hours on an average to search for the user's Facebook identity. However, the huge search time involved sleep time, because of the Twitter API rate limit (350 requests / hr). Network search method involved searching for all the users in nemo's network, who self-identified themselves. For the same, the method queried each user in nemo's network composed of thousands of users (767 followers for the user identified). During the search, the rate limit exceeded, and the method had to wait till the next hour. We optimized the algorithm to do processing in the mean time, as well as quit as soon as two users in the network find nemo's Facebook identity, however we could reduce the time to 1.4 hours for the user identified. Network Search algorithm gives the correct Facebook identity for a small set of users, within a minimum candidate set size, however takes huge search time.\\
\indent With each of the algorithms, we observe that as we experiment with attributes ranging from generic attributes to specific attributes about a user, candidate size decreases and becomes more precise however at the cost of more time and less accuracy. The reason is that more specific attributes expose precise information about the user, which is true for small set of users, while generic attributes expose information about a user, which can be true for large set users also. However, computing precise information takes time and therefore the search time is the highest for search via network. Owing to the variation of accuracy of identity search algorithms,  we integrated all the four methods together to build a system, which must capture available identity exposures via any dimension -- profile, content and network. 
%\begin{figure}[ht]
%  \centering
%  \includegraphics[scale=0.25]{idea.pdf}
%  \caption{Figure shows the trend as we move from exploiting generic to specific attributes of a user to search for her identity. Possible candidate set size decreases while search time increases. }
%  \label{fig:idea}
%\end{figure}
\vspace{-3mm}
\subsection{Integrated System}
\indent Owing to low accuracy of each method, we built an integrated architecture by combining all four algorithms in one system named as ``Finding Nemo".~\footnote{http://precog.iiitd.edu.in/findingnemo/home} The system can be queried with a Twitter identity to search her Facebook identity. Any of the four algorithms can locate her Facebook identity. If a candidate identity is returned by either self-identification, self-mention or by more than one algorithm, the candidate identity is returned by the system as the Facebook identity of the user. In other scenarios, candidate identities returned by each algorithm are ORed. The system returns a set of Facebook identities as which may / may not contain correct Facebook identity.\vspace{-3mm}
%\begin{figure}[ht]
%  \centering
%  \includegraphics[scale=0.2]{nemo.pdf}
%  \caption{Architecture of an online identity search system -- Finding Nemo. }
%  \label{fig:nemo}
%\end{figure}
\subsubsection{Accuracy}
We measured the effectiveness of Finding Nemo by evaluating the system for the same dataset of 543 users. We observed that for 220 Twitter users (40.5\%), correct Facebook identity were identified within the candidate identities returned by the system. Table~\ref{tab:all} lists the accuracy of individual algorithms as well as of the complete integrated system. \vspace{-3mm}
 \begin{table}[h]
   \centering
   %\topcaption{Table captions are better up top} % requires the topcapt package
   \begin{tabular}{|l|p{2cm}|p{2cm}|} % Column formatting, @{} suppresses leading/trailing space
     \hline
      \textbf{Search Algorithm}   & \raggedright{ \textbf{\# of users identified}} & \textbf{Accuracy} \\ \hline
          Profile Search (P) & 205 & 37.7\%\\ \hline
          Content Search (C) &  3 & 0.5\% \\ \hline
          Self-mention Search (SM) & 31 & 5.7\% \\ \hline
          Network Search (N) & 1 & 0.2\% \\ \hline
          Finding Nemo & 220 & \textbf{40.5\%} \\ \hline
   \end{tabular}
   \caption{Accuracy of each algorithm, and the system Finding Nemo.}
   \label{tab:all}
\end{table}
\vspace{-3mm}
\subsubsection{Candidate set size}
Figure~\ref{fig:n_cand} shows the candidate set size distribution for the users who were correctly identified on Facebook by the system. We observed that for majority of users, candidate set size largely reduced while for some increased, as compared to individual algorithms. The reason being that for majority of users, their Facebook identity was exposed either by self-identification, self-mention and by more than one method. As soon as the system encountered a user's Facebook identity exposure in any of the three scenarios, it returned the Facebook identity, without processing other methods for identity search. However, for other users, no candidate Facebook identity existed which was found by more than one method and therefore, system collated all candidate identities returned by individual methods, thereby increased the candidate set size. \vspace{-3mm}
\subsubsection{Search time}
\begin{figure}[htp] %  figure placement: here, top, bottom, or page

              \subfigure[pari\_lakshya]{
     
     \fbox{    \includegraphics[scale=0.2]{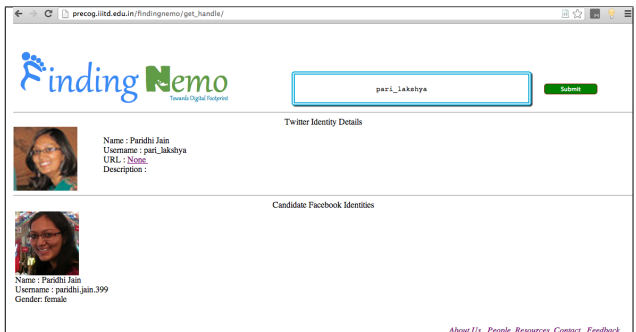}}
         \label{fig:esimperl}
    }     
%      \subfigure[hindujagrutiorg]{
%      
%       \fbox{  \includegraphics[scale=0.2]{hindujagrutiorg.pdf}}
%         \label{fig:hindujagruti}
%    }
         \vspace{-4mm}
    \caption{Finding Nemo: screenshot}
    \vspace{-4mm}
    \label{fig:screenshot}
 \end{figure}

We analyzed search time of the system to search for users' Facebook identity in real-time. We queried system at different times of the day to avoid Twitter rate limit expiry. Figure~\ref{fig:nemo_time} shows the search time distribution for the users correctly identified by the system. The system came to halt as soon as a Facebook identity was recognized via self-identification, self-mention or by more than one method. Therefore, majority of users were identified within 50 secs (system did not need to exploit all four methods). For one user, identified by network search method and no other method, took 5,535 secs and therefore system took too much time searching via all the four methods.~\footnote{System processed four algorithms in a serial manner -- Profile, Self-mention, Content, and Network. } \\
\indent In conclusion, Finding Nemo returned reduced candidate size to search in for majority of users, however at the cost of time. Figure~\ref{fig:screenshot} shows screenshot of Finding Nemo when searched for Twitter user - pari\_lakshya. pari\_lakshya exposed her Facebook identity via her profile attributes as well as her network attributes. The system is successful in finding groups' and celebrities' Facebook identity also. We evaluated each identity search algorithm on a test bed of two most popular social networks and observed a non-zero accuracy of each algorithm. We therefore emphasize on the possibility of success of the algorithms when applied with private information access (apart from publicly available), and on other social networks.
\setlength\fboxsep{0pt}
\setlength\fboxrule{0.3pt}
\vspace{-3mm}
\section{Discussion} \label{Discussion}
In this work, we explore the possibility of searching a user identity on multiple social networks, by exploiting public information about her. A user gives away references to her identities in her publicly accessible profile, content and network attributes, knowingly or unknowingly. Even if she consciously does not leave any identity leaks, her identity can be leaked via loose privacy decisions of her friends, thereby posing a threat to her privacy. We attempt to exploit only those allusions to a user's social network identity, which are supported by the involved social networks (Twitter and Facebook). The algorithms need to be reformulated if they are applied on other social networks depending on the information support a social network provides. For example, LinkedIn API does not provide access to the network of any user unless the user authorizes via an app. Therefore, a random user's identity search via network search algorithm is not applicable without the user's authorization. Applicability of each algorithm is dependent on the information support infrastructure provided by the social networks on which a user's identity is known and searched.\\
\indent Further, the effectiveness of each algorithm and the system depends on the presence of the required identity information provided by the user's attributes on the considered dimension. The algorithm fails to find other social network identity for a user, if a user is self-aware and informed about certain identity exposures (e.g. no mention of urls in tweets or restricted access to friends). This is evident by different accuracy achieved by each algorithm. However, we successfully reduce the candidate set of identities to search in, thereby saving time and effort. \\
\indent Apart from Facebook, we observe that users often self-mention their identity on few photo sharing and video sharing social networks, via URLs posted in tweets pointing to the pictures / videos uploaded on the networks. Table~\ref{tab:booktabs3} shows the ranked list of the social networks embedded in URL by 2132 Twitter users (randomly selected from bigger dataset). With multiple exposed identities of a user, a detailed footprint can be created by aggregating user's details from Twitter, Facebook, other networks which may lead to exposure of certain private attributes e.g. gender, date of birth, family etc.  \\
 \begin{table}[htbp]
   \centering
   %\topcaption{Table captions are better up top} % requires the topcapt package
   \begin{tabular}{|c|p{3cm}|} % Column formatting, @{} suppresses leading/trailing space
     \hline
	\raggedright { \textbf{Social Network } }   & \textbf{ \% of users } \\ \hline
          Instagram & 36.6\% \\ \hline
          Youtube & 29.7\% \\ \hline
          Foursquare & 6.1\% \\ \hline
          Tumblr & 6.0\% \\ \hline
     	Yfrog  & 4.0\% \\ \hline 
   \end{tabular}
   \caption{Social networks embedded in the URLs.}
   \label{tab:booktabs3}
\end{table}
\indent We argue that systems like Finding Nemo can be used to help an analyst to search for malicious user identities, banned organizations, a friend to search for a friend's social network identities, and social web administrator to build a footprint of a user to disallow creation of fake identities, given the support of social networks. However, we also envision that such systems can be exploited to breach a user's privacy and to target victims for other attacks e.g. spear phishing.
\vspace{-2mm}
\subsection{Limitations} \label{Limitations}
\subsubsection{Algorithms}
The success of the proposed algorithms is limited by variety of factors.  Firstly, algorithms are dependent on the exposure of required information by a user and on the social network information support. For each social network, algorithms need to be reformulated, to include other hints of identity exposures available. Secondly, the algorithms exploit only few identity leaks and therefore fail if the discussed identity leaks are patched. The algorithms are not scalable to find other hints for identity search. Thirdly, algorithms are highly dependent on a social network support for a better recall and lastly, few algorithms (Self-mention and Network search) have large search time owing the social network restrictions on number of queries, which may not be useful when there is a need to find a user's multiple identities in runtime fashion.
\vspace{-3mm}
\subsubsection{Evaluation}
We choose a random set of 543 users out of 505,466 users, which may include users who had certain identity leaks leading to better accuracy for a particular method and low accuracy for other algorithms. As discussed, the results are highly dependent on the quality of identity exposure provided by the user and therefore the dataset used for evaluation. The results can be biased by the dataset used for evaluation, however we try to minimize the bias by using a random sample from a bigger dataset. We plan to evaluate the algorithms and system over a large dataset of users, specifically targeting various age groups, profession and demographics, to claim the generic success of the algorithms and therefore the success of Finding Nemo.
\vspace{-3mm}
\subsection{Future Work} \label{Conclusion}
Our system, Finding Nemo successfully found Facebook identities for 40\% of the users (219) given their identity on Twitter. To find the other 60\% of the users, the algorithms can be re-devised and improved to more sophisticated approaches to capture granularities present in identity leaks e.g authorship profiles, temporal characteristics, timing profiles and network characteristics. We plan to extend Finding Nemo to search for a user's identity on any social network in real-time, given her identity on any social network in future. We also envision the system to provide feedback to the user about her possible identity exposures and suggest to patch them, in order to preserve privacy across networks.
%\section{Acknowledgment}
%The authors would like to thank TCS Research Fellowship Program for their support, and all members of PreCog research group at IIIT-Delhi for their valuable feedback and suggestions. Special thanks to Anshu Malhotra and Luam Totti for sharing the dataset and Siddhartha Asthana for his feedback during the development of this paper.
%\subsection{}
\vspace{-3mm}
{
\bibliographystyle{IEEEtran}
\bibliography{references1}
}

\end{document}